# Mn clusterisation in $Ga_{1-x}Mn_xN$

D. Wang[1], X. Y. Zhang[1,*], J. Wang[2,3], S. Q. Wei[2], W. S. Yan[2], and D. W. Boukhvalov[4]

1. Department of Physics, Surface Physics Laboratory (National Key Laboratory), and Synchrotron Radiation Research Center, Fudan University, 220 Handan Road, Shanghai 200433, China
2. National Synchrotron Radiation Laboratory, University of Science and Technology of China, Hefei 230029, China
3. Shanghai Synchrotron Radiation Facility, Shanghai Institute of Applied Physics, Shanghai 201800, China
4. Institute for Molecules and Materials, Radboud University of Nijmegen,
   NL-6525 ED Nijmegen, the Netherlands

**Abstract**

Local structure of Mn atoms in $Ga_{1-x}Mn_xN$ has been investigated by the Mn $L_3$ edge x-ray absorption spectrum (XAS) at total electron yield mode, which preferentially looks at atoms near the surface. A modeling defects configuration, the $Mn_5$ micro-clusters complexed with substitutional $Mn_{Ga}$ and interstitial $Mn_I$ is found for the higher Mn doping concentration. This new configuration is also confirmed by the total energy calculations.

Corresponding author, email address: xy-zhang@fudan.edu.cn

# 1. Introduction

The combination of electronic and magnetic properties in a single device opens many prospects for information technology [1]. To this end, the unique properties are found in the III-V diluted magnetic semiconductors (DMSs), in which the long range magnetic ordering between localized d-electrons of the magnetic impurities is thought to be mediated by an interaction with the itinerant valence electrons of the host semiconductor. Interests are further caused by the prediction of ferromagnetism at room temperature and even above in these materials [2-4]. Of the established DMSs, Mn-doped GaN films are predicted to have the higher Curie temperatures ($T_C$) above room temperature [5] and many efforts involving the controlled fabrication of these materials as well as to explore the characterization and their physical properties have been developed [6-8]. It has been suggested that the ferromagnetism (FM) in (Ga,Mn)N could arise from the formation of magnetic clusters [9]. Kane *et al.* reported that the room temperature ferromagnetic hysteresis was observed in epitaxial layers of $Ga_{1-x}Mn_xN$ with concentrations of up to $x = 0.015$ grown on c-sapphire substrates by metalorganic chemical vapour deposition, which might be due to either Mn-clustering on the atomic scale or the $Ga_{1-x}Mn_xN$ bulk alloy [10]. Zając *et al.* pointed out that different $Mn_xN_y$ phases were responsible for ferromagnetic behavior of (Ga,Mn)N grown by the ammonothermal and chemical transport methods, and the $Mn_xN_y$ phases could be either in the form of tiny clusters or they were of small volume. They thought that some of the $Mn_xN_y$ phases were ferro/ferrimagnetic and thus provided (high-temperature) FM contribution to the total magnetization of (Ga,Mn)N samples [11]. However, K. Biswas *et al.* have prepared Mn-doped GaN nanocrystals under solvothermal conditions by keeping the temperature of the preparation around 350 °C.

All the nanocrystals show ferromagnetism at room temperature, and the secondary phases are unlikely to be formed under the preparative conditions employed [12]. Although a great deal of emphases has been placed on understanding the origin of the ferromagnetic behavior of GaMnN, it is still in debate. One of the reasons is the complexity and instability of DMS materials in the preparation processes, which are frequently affected by some untraceable factors. Recently, we studied the 3$d$ electronic states of Mn dopants in $Ga_{1-x}Mn_xN$ by Mn $L$-edge X-ray absorption spectroscopy (XAS) measurements. We have shown that most of Mn atoms in $Ga_{1-x}Mn_xN$ are bivalent ($Mn^{2+}$) with the total spin $S=5/2$ and the $p$-$d$ hybridization is strongly influenced by Mn doping concentrations. When x□0.025□the $Ga_{1-x}Mn_xN$ with stronger $p$-$d$ hybridization is obtained [6]. In addition, our previous near edge x-ray absorption fine structure (NEXAFS) study in Mn $K$ edge [13] shows that most Mn atoms locate at Ga substitutional sites, $Mn_{Ga}$ at low Mn content and the $Mn_4$ clusters, being made of four Mn atoms, exist at the higher doping concentration. Edmonds *et al.* [14] found that the interstitial $Mn_I$ was relatively mobile and had a tendency to outdiffuse to the surface. It is not unreasonable to expect that the local structure of Mn atom should be different in bulk or near surface. The clusterisation of Mn in GaMnN might play an important role in magnetic properties and perhaps results in the non-reproducibility of good quality DMSs, especially when the grain sizes of GaMnN are small. The traditional method to study the different local structures in films in near surface layer is using the etching [14]. It would destruct the surface of the film and does not suit for thin films. In this paper, we investigate the Mn cluster structures in $Ga_{1-x}Mn_xN$ by using a nondestructive method based on the x-ray absorption studies at Mn $L_3$ edge.

**2. Experimental**

As we know, the x-ray penetration length is the reciprocal of the absorption coefficient. The x-ray penetration depths for the Mn $K$ edge (6540 eV) and $L$ edge (640 eV) incident energies are 3 μm and 70 nm, respectively. While the Mn $L$ edge measurement is performed in total electron yield mode, only electrons that come from the sample within 50 Å of the surface layer, which is the electron escape depth, will be collected. Then, the escape depth of x-ray photons is about 1μm. Therefore, the local structure of Mn in bulk can be studied by the $K$ edge NEXAFS at its x-ray fluorescence mode [13], and the structure of atoms in the near surface layer can be monitored by using the $L$ edge x-ray absorption in total electron yield mode.

Two cubic $Ga_{1-x}Mn_xN$ films about 0.3 μm thick with different Mn concentrations were prepared by the plasma-assisted molecular beam epitaxy (PA-MBE) on GaAs (001) substrates at the University of Nottingham [15]. The growth was performed under nitrogen-rich conditions and the substrate temperature was fixed at 140 ºC. A GaAs buffer layer (about 0.15 μm thick) was first grown on the substrate to improve the quality of the $Ga_{1-x}Mn_xN$ layer. The concentrations of Mn were determined by the secondary ion mass spectrometry. Samples with x=2.5 at.% and 10 at.% were used in this work. The x-ray diffraction (XRD) measurements confirmed the structure of a zinc-blende type for both films.

The x-ray absorption spectra were recorded at room temperature in total electron yield mode at the beamline U18 of the National Synchrotron Radiation Laboratory (NSRL). The beamline covered an energy range from 100 to 1000 eV with a resolution of about 0.2 eV. The photon incident angle was set to be 50º to the surface normal. The Raman experiments were carried out in the backscattering geometry at room temperature using the 514.5 nm line of an Ar-ion laser as the excitation source.

**3. Results and discussion**

Figure 1 shows the experimental Mn $L_3$ edge XAS of $Ga_{1-x}Mn_xN$ thin films with two different Mn doping concentrations (x=2.5% and x=10%, respectively). Both spectra have been subtracted by the pre-edge background, and then normalized by the incident intensity $I_0$ and the maximum of the peak heights at the absorption edge. For the sample of x=2.5%, the $L_3$ edge exhibits one peak feature (labeled A) accompanied by a very small shoulder (labeled B) at the higher energy side. While for the higher Mn concentration (x=10%), the peak A becomes narrower and two additional peaks (labeled C and D) appear obviously in the high energy side of peak A. This profile is a specific characteristic of the highly localized $d^5$ ground state [16]. Y. Yonamoto *et al.* [17] and Y. Ishiwata *et al.* [18] thought that the large multiplet splitting feature resulted from the Mn oxide at surfaces due to highly sensitive of Mn $L_3$ edge spectra to the surface in total electron yield mode. Edmonds *et al.* [19] pointed out that the Mn oxide in the surface was less of an issue in the material (Ga,Mn)N than in (Ga,Mn)As. A shift of about 0.5 eV to higher energies should be observed in $L_3$ spectra after Mn oxidizing. However, no such an energy shift was observed in XAS spectra of our two samples. It seems that the large multiplet splitting feature in the 10% Mn doping concentration sample comes from Mn atoms but not from the surface oxide contamination.

As we know, various configurations of different Mn occupations associated with diverse kinds of adjacent defects will affect the local electronic structure of Mn atoms, which corresponds to different features in Mn $L_3$ edge XAS. For understanding the nature of Mn in zinc-blende $Ga_{1-x}Mn_xN$, the Mn $L_3$ edge XAS curves were calculated by FEFF 8.2 code [20]. Different representative models of the $Ga_{1-x}Mn_xN$ with various Mn occupation sites in GaN lattice are used for the XAS calculations. Several typical ones are shown in Fig. 2. They are substitutional Mn ($Mn_{Ga}$), interstitial Mn ($Mn_I$), $Mn_4$ micro-cluster (four congregated $Mn_{Ga}$ neighbors) and $Mn_5$ micro-cluster (four neighbors of Mn$_{Ga}$ adding one Mn$_I$ in the center). The XAS calculations of Mn$_{Ga}$ or Mn$_I$ with several possible defects as neighbors are also considered. The $L_3$ edge XAS calculations used the Ga$_{1-x}$Mn$_x$N cluster with 123 atoms and carried out using initial ground-state potentials and a fully relaxed hole at the core level. The Dirac-Hara model of exchange potential with a 1.2 eV shift and an additional broadening of 0.1 eV is used for the $L_3$ edge XAS calculation. All calculations are done using the theoretical lattice constant (a= 4.46 Å).

Figures 3(a) and 3(b) show the calculated $L_3$ edge spectra using different Mn configurations as mentioned above. It can be seen that none of the calculated XAS spectra with one single type of Mn configuration can well fit the experimental results. Therefore, numbers of simulations based on various combinations of different Mn configurations are performed. The simulation curve of one substitutional Mn$_{Ga}$ plus a Mn$_{Ga}$ bound to an N vacancy nearby at a ratio of 1:1 reproduces most closely the experimental line shape of the sample at 2.5% Mn doping, as shown as curve *b* in Fig. 4, and the appearance of a shoulder B in curve *a* is mostly due to the N vacancy bound to the Mn atom nearby (cf. Fig. 3(b)). While for 10% Mn doped sample, the simulation is more complicated due to the narrower full width at half maximum (FWHM) as well as the appearance of a new peak D at higher energy side. The model, including the Mn$_4$ micro-cluster, Mn$_I$ and Mn$_{Ga,}$ used in our previous NEXAFS study for Mn *K* edge [13], was first tested. Unfortunately, it could not give a satisfactory fitting. Then, as many as possible combined models were constructed. Some models that were more complicated than the one suitable for *K* edge XAS of GaMnN were used. The results show that a model made of Mn$_5$, Mn$_{Ga}$, and Mn$_I$ is the most probable one. The simulation curve, a linear addition of calculated spectra of Mn$_5$, Mn$_I$, and Mn$_{Ga}$ in ratio of 2:2:1 approximately, is well consistent with our experimental data, as shown as curve *d* in

Fig. 4. It surprised us that for two cases (*K* edge and *L₃* edge), the only difference exists between Mn$_4$ and Mn$_5$ micro-clusters.

To verify the possibility of forming Mn$_5$ micro-clusters, we carried out *ab initio* calculations. For our calculations we used SIESTA pseudopotential code [21, 22], which was before implemented to calculation of DMS [23]. In our calculation, we studied the changes of formation energy per Mn atom for super-cell contained 32 atoms of Ga and 32 atoms of N. The model concentration of Mn 15.6% is close to the experimental one. Calculations of formation energies were performed by standard formulas used in [23]. For verifying our method and technical parameters, we carried out the LDA [24] calculation of pure GaN with full optimization of lattice parameters and atomic positions. A small (less than 1%) compressing of lattice parameter from experimental one was observed.

References [25, 26] suggest us for correct calculation of formation energy, which is needed to find optimal magnetic configurations. For calculating the ground state of Mn, we performed the calculation for antiferromagnetic (AFM) state of γ-Mn. For pair of substitutional Mn in GaN, we found that the AFM configuration had a lower energy than ferromagnetic one. For this configuration the formation energy was about 1.69 eV per Mn, which was a little smaller than one for single substitutional Mn (1.77 eV) and compressing of lattice parameter by about 1.5% from experimental one. Next, an increase of Mn micro-clusters, which contained only substitutional Mn atoms, provided a neglected decreasing of formation energy per Mn to 1.65 eV for Mn$_4$ micro-cluster presented in Fig. 2(c). Next step in our modeling of Mn impurities in GaN was to study the probability of forming micro-clusters, which contained substitutional and interstitial atoms. The possibility of substitutional and interstitial dopants and their connection with ferromagnetism in DMS thin films were discussed in Refs. [23, 27]. First we calculated the formation energy per single interstitial Mn, which was 7.60 eV. However, the combination of substitutional and interstitial Mn provided a strong decreasing formation energy to 3.25 eV per Mn. We need to note that for this type of configuration, lower energy should correspond with the AFM configuration, which could be explained by the presence of a strong direct AFM Mn-Mn exchange as in pure γ-Mn. It is similar to the results reported recently in Ref. [28] about AFM coupling between substitutional and interstitial Mn. Then, an increase of the size of micro-clusters provided a decreasing of formation energy per Mn to 1.55 eV for $Mn_5$ micro-cluster, which contained four substitutional and one interstitial Mn atoms (see Fig. 2(d)). Compressing of lattice parameter for described $Mn_5$ micro-cluster is 0.4% from experimental one for pure GaN. Therefore, we can summarize that described micro-cluster is most favorable type of impurity in studied samples.

Based on our simulation above, it is concluded that the Mn atoms in the film is mostly $Mn_{Ga}$ substitutions for 2.5% Mn doping concentration, and for 10% Mn doping concentration, in addition to substitutional $Mn_{Ga}$ and interstitial $Mn_I$, more than 75% Mn atoms aggregate into $Mn_5$ micro-clusters. The formation of $Mn_5$ micro-clusters depends on the concentration of both $Mn_{Ga}$ and $Mn_I$. The outdiffused $Mn_I$ might accumulate at the surfaces of the $Ga_{1-x}Mn_xN$ film leading to not only the more possibility of the formation of $Mn_5$ micro-cluster, but also further decreasing the $Mn_I$ in the bulk of the film. Consequently, the Mn atoms near the surface intend to be aggregated together in the form of $Mn_5$ micro-clusters, while $Mn_4$ micro-clusters in the bulk.

To obtain additional information about the incorporation of Mn ions into the GaN lattice, the Raman spectra of our samples are presented in Fig. 5. For comparing with our doping samples, the Raman spectra of the pure cubic GaN and the GaAs are also given. The two peaks located at 550 cm$^{-1}$ and 736 cm$^{-1}$ are observed for both samples. They are the characteristic features of the cubic GaN (LO and TO vibrational modes) and also appear in the Raman spectrum of the pure cubic GaN, as shown in Fig. 5. The broader and asymmetric LO band means the lower electron concentration [29] due to the existence of defects and Mn$_I$ in our films. Two peaks around 667 cm$^{-1}$ and 586 cm$^{-1}$ in the spectra of Mn doped samples are disorder-activated modes or local vibrational mode (LVM) of the GaN:Mn [30] and marked with "a" and "b" in the figure, respectively. These modes can be observed when long-range lattice ordering of the host lattice is lost. A relatively strong peak "a" is assigned to a LVM of GaN related to a N vacancy [31]. The peak "b" is assigned to the LVM of Mn occupying the Ga site. When the doping concentration is increased, the decreasing intensity of peak "b" means that the number of the substitutional Mn$_{Ga}$ decreases. It is well consistent with above $L_3$ edge XAS studies.

## 4. Conclusion

In summary, we have shown that the Mn occupations in GaN are strongly influenced by the Mn doping concentrations and have different geometry configurations in near surface layer or in the bulk. Mn atoms near surfaces in Ga$_{1-x}$Mn$_x$N can be directly probed by the Mn $L_3$ edge XAS. A new structural model of Mn$_5$ micro-clusters is proposed for the GaMnN surface layer, which explains well the $L_3$ edge XAS experimental results, and is also supported by our total energy calculations. The combined analysis of Mn $L_3$ edge and $K$ edge XAS measurements is applied to identify the different clusterisation of Mn atoms near surface or in bulk and provides a useful method for understanding the cluster formation of Mn in GaMnN thin films.

**Acknowledgements**

This work was supported by the Project of the State Key Program of National Natural Science Foundation of China (Grant No. 10635060). The authors are very grateful to C. Thomas Foxon and Dr. S. V. Nivikov of the University of Nottingham for providing the $Ga_{1-x}Mn_xN$ samples, to Kazutoshi Mamiya of PF and Honghong Li of the NSRL for the help in XAS measurements, also to Yan Zhu and Zhongqin Yang for valuable discussions in theoretical calculations. DWB acknowledges financial support by Stichting voor Fundamenteel Onderzoek der Materie (FOM), the Netherlands.

**Figures**

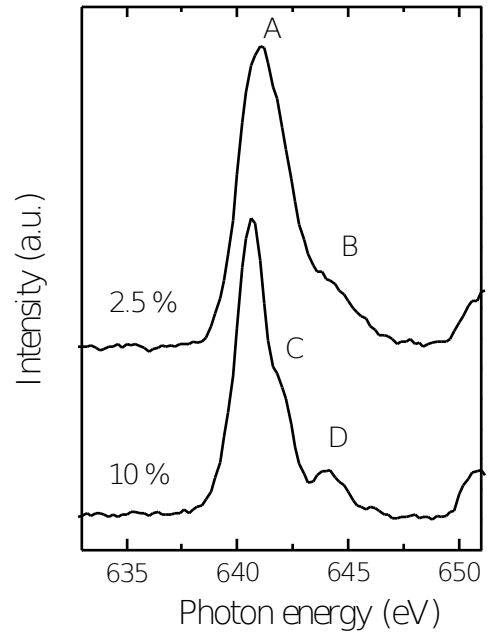

Fig.1 Mn $L_3$ edge absorption spectra of $Ga_{1-x}Mn_xN$ thin films.

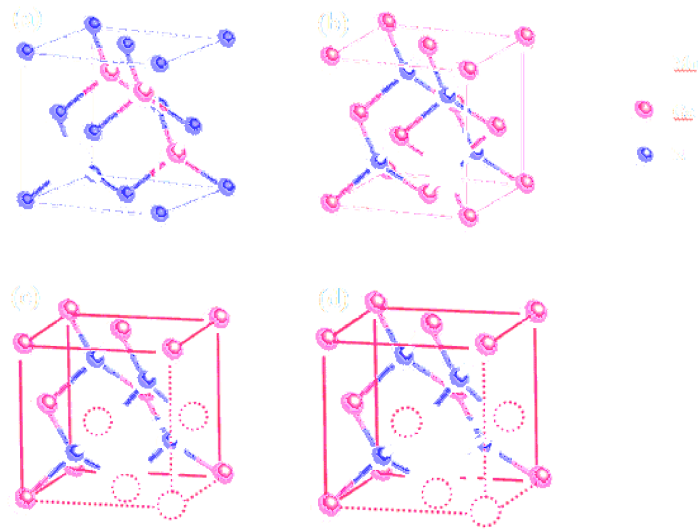

Fig.2 (Colour online) Model structures of various Mn occupation sites in GaN lattice: (a) substitutional $Mn_{Ga}$, (b) interstitial $Mn_I$, (c) $Mn_4$ micro-cluster and (d) $Mn_5$ micro-cluster. The empty circles with dotted-line border represent the locations in the zinc-blende structure.

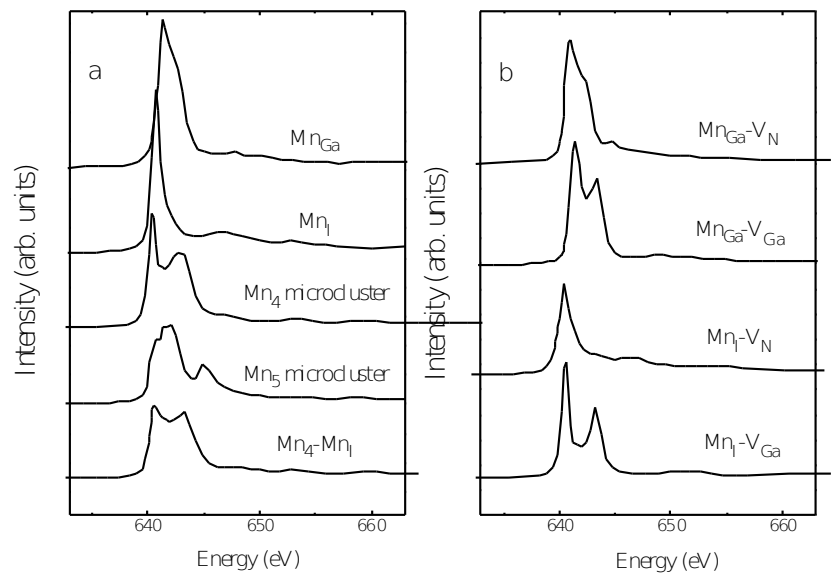

Fig.3 Calculated $L_3$ edge XAS spectra (a) for various Mn configurations in GaN lattice, and (b) with various related defects near the Mn atom in the film.

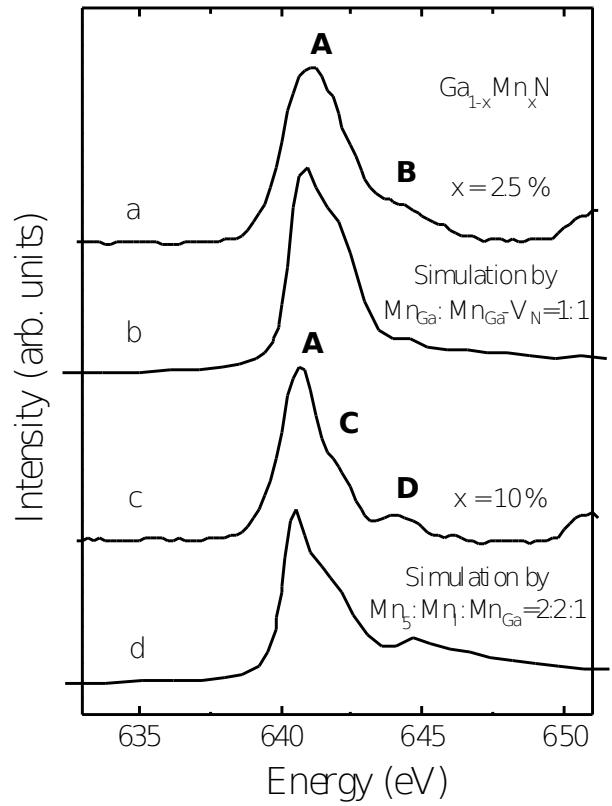

Fig.4 Comparison of the Mn $L_3$ edge XAS spectra from theoretical calculations and experimental spectra for $Ga_{1-x}Mn_xN$ with various Mn concentrations. Curves *a* and *c* are the experimental spectra for x=2.5% and 10%, respectively, and *b* and *d*, the simulation curves for both samples respectively.

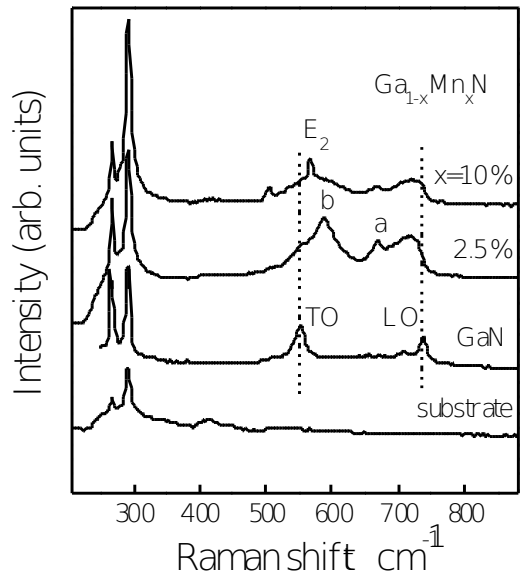

Fig.5 Raman spectra of $Ga_{1-x}Mn_xN$ films.